\documentstyle[sprocl,epsfig]{article}

\def\be{\begin{equation}}
\def\ee{\end{equation}}
\def\bea{\begin{eqnarray}}
\def\eea{\end{eqnarray}}

\newcommand{\AmS}{{\protect\the\textfont2
  A\kern-.1667em\lower.5ex\hbox{M}\kern-.125emS}}
\newcommand{\lsim}{\raisebox{-3pt}{$\,\stackrel{\textstyle <}{\sim}\,$}}
\newcommand{\gsim}{\raisebox{-3pt}{$\,\stackrel{\textstyle >}{\sim}\,$}}

\def\Pom{{\bf I\!P}}

\def\lsim{\mathrel{\rlap{\lower4pt\hbox{\hskip1pt$\sim$}}
    \raise1pt\hbox{$<$}}}         

\def\gsim{\mathrel{\rlap{\lower4pt\hbox{\hskip1pt$\sim$}}
    \raise1pt\hbox{$>$}}}         

\begin{document}

\title{
SPIN DEPENDENCE OF DIFFRACTIVE VECTOR MESON PRODUCTION}



\author{N.N. Nikolaev}

\address{Institut f. Kernphysik, Forschungszentrum J\"ulich,
D-52450 J\"ulich, Germany\\
 L.D.Landau Institute for Theoretical Physics,
142432 Chernogolovka, Russia
\\E-mail: N.Nikolaev@fz-juelich.de } 


\maketitle\abstracts{ 
I review the recent progress in the theory of $s$-channel helicity 
nonconservation (SCHNC) effects in diffractive DIS. SCHNC in
diffractive vector meson production is an unique probe of
the spin-orbit coupling and Fermi motion of quarks in
vector mesons. Photo- and electroproduction of the $\phi$ and its
radial and angular excitations at Jlab is
perfectly posed to probe SCHNC in QCD pomeron exchange.
I also discuss the unitarity driven demise of 
 the Burkhardt-Cottingham sum rule and large \& scaling departure from 
the Wandzura-Wilczek relation.}

\section{Introduction}

The common wisdom is that spin-dependence vanishes in the high energy
limit, as an example recall the rapidly decreasing longitudinal
spin asymmetry , $A_{1}$, in polarized DIS. 
In high energy QCD the well known quark helicity 
conservation  was
believed universally to entail the $s$-channel helicity conservation (SCHC)  
at small $x$, i.e. the QCD pomeron was supposed to decouple from
helicity flip, i.e., spin-dependence of diffractive DIS 
was supposed to vanish st small $x$. Here I review the recent work 
\cite{NZDIS97,NPZLT,KNZ98,IN99,AINP,G2,IK},  which shows this belief 
was groundless, and expose an extremely rich SCHNC physics 
in diffractive DIS at small $x$. Furthermore, by the virtue
of unitarity diffractive SCHNC 
is found to change dramatically the small-$x$ behaviour 
of transverse spin asymmetry $A_{2}$
and lead to the demise of the Burkhardt-Cottingham sum rule and 
the departure from the Wandzura-Wilczek relation.

\section{Is SCHNC compatible with quark helicity conservation?}

The backbone of DIS  is the Compton scattering (CS) $\gamma^{*}_{\mu}p\to 
\gamma^{*'}_{\nu}p'$,
which at small-$x$ can be viewed as a (i) dissociation 
$\gamma^{*}\to q\bar{q}$ 
followed by (ii) elastic scattering $q\bar{q} p \to q\bar{q} p'$ 
with {\bf exact}
conservation of quark helicity and (iii) fusion $q\bar{q} \to \gamma^{*'}$. 
The CS amplitude $A_{\nu\mu}$ can be written as 
$
A_{\nu\mu}=\Psi^{*}_{\nu,\lambda\bar{\lambda}}\otimes A_{q\bar{q}}\otimes
\Psi_{\mu,\lambda\bar{\lambda}}
$
where $\lambda,\bar{\lambda}$ stands for $q,\bar{q}$ helicities,
$\Psi_{\mu,\lambda\bar{\lambda}}$ is the wave function of the $q\bar{q}$ 
Fock state of the photon. The $q\bar{q}$-proton scattering
kernel $A_{q\bar{q}}$  does not depend on, and conserves exactly, 
the $q,\bar{q}$ helicities. For
nonrelativistic massive quarks, $m_{f}^{2} \gg Q^{2}$, one only has
transitions $\gamma^{*}_{\mu} \to q_{\lambda} +\bar{q}_{\bar{\lambda}}$ 
with $\lambda +\bar{\lambda}=\mu$. However, 
the relativistic P-waves give rise to 
transitions
of transverse photons $\gamma^{*}_{\pm}$ into the $q\bar{q}$ state with 
$\lambda +\bar{\lambda}=0$
in which the helicity of the photon is transferred to the $q\bar{q}$ orbital 
angular momentum. Consequently, the SCHNC transitions
$\gamma^{*}_{\pm} \to (q\bar{q})_{\lambda +\bar{\lambda}=0} \to
\gamma^{*}_{L} ~~~{\rm and}~~~
\gamma^{*}_{\pm} \to (q\bar{q})_{\lambda +\bar{\lambda}=0 }\to
\gamma^{*}_{\mp} $ 
are allowed \cite{NZ92,NZDIS97,NPZLT}. Furthermore, the SCHNC and SCHC amplitudes,
which were first calculated in \cite{NZ92} and called ${\bf \Phi}_{1}$ and
${\bf \Phi}_{2}$ there, respectively, 
have similar $x$-dependence, i.e., the QCD pomeron exchange contributes to
the both SCHC and SCHNC transitions. We emphasize that the above argument 
for SCHNC does not require applicability of pQCD.

\begin{figure}
\epsfysize 4 in 
\epsfbox{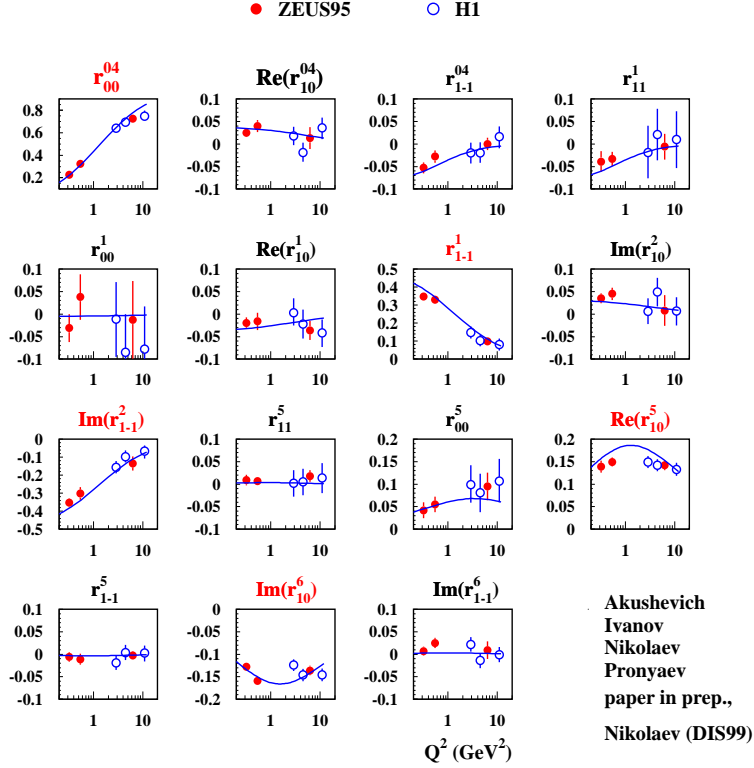}
\caption{Our prediction \protect\cite{AINP} 
for the 
spin density matrix $r_{ik}^{n}$ 
of diffractive $\rho^{0}$-meson vs. the experimental
data from
ZEUS \protect\cite{ZEUS} and H1 \protect\cite{H1}.}
\label{fig:1}
\end{figure}

By conservation of the angular momentum the single-flip $T\to L$ and double-flip
$\pm \to \mp$ transition amplitudes are $\propto \Delta$ and 
$\propto \Delta^{2}$, respectively. Here ${\bf \Delta}$ is the
$(\gamma,\gamma')$ and/or $(p',p)$ momentum transfer.

\section{LT interference in diffractive DIS into continuum
$\gamma^{*}p \to p'X$}

The first ever direct evaluation \cite{NZDIS97} of SCHNC effect in QCD - the
 LT-interference 
of transitions $\gamma^{*}_{L}p \to p'X$ and $
\gamma^{*}_{\pm}p \to p'X$ into the same continuum 
diffractive states $X$  - has been reported in 1997 and
went unnoticed. Experimentally, it can be measured at HERA by both H1 and
ZEUS via azimuthal correlation between the $(e,e')$ and 
$(p,p')$ scattering planes. The detailed discussion of this  
asymmetry $A_{LT}$ and its use for the determination of the otherwise elusive 
$R=\sigma_{L}/\sigma_{T}$ for diffractive DIS is found in \cite{NPZLT}. 
Here I only
recall that azimuthal asymmetry is the twist-3 effect,
\begin{equation}
A_{LT} \propto {\Delta \over Q}g_{LT}^{D}(x_{\Pom},\beta,Q^{2})\, ,
\label{eq:2}
\end{equation}
where $g_{LT}^{D}$ is the scaling structure function. It does not decrease
at $x\to 0$!

\section{SCHNC in diffractive $\gamma^{*}p \to Vp'$}

Evidently,  diffractive 
vector meson production $\gamma^{*}p\to Vp'$ is obtained from CS
 by continuation from spacelike $\gamma^{*}$ to timelike 
$V$. The azimuthal correlation
of the $(e,e')$ and $(p,V)$ planes with the vector meson decay 
plane, and the in-decay-plane
angular distributions of decay products, allow the experimental
determination of all helicity amplitudes $A_{\mu\nu}$. 
The consistent analysis of the 
$S$-wave and $D$-wave states of the vector mesons is presented only
in \cite{IN99}. At small $x$ and within the diffraction cone helicity
amplitudes for 
$\gamma^*_{\mu} \to V_{\nu}$ have the following
form \cite{KNZ98,IN99}
\begin{eqnarray}
A_{\nu\mu}(\bar{x},Q^{2},{\bf \Delta})=
is{c_{V}\sqrt{4\pi\alpha_{em}}
\over 2\pi^{2}}\exp(-{1\over 2}B_{3\Pom}\Delta^2)
\int_{0}^{1} {dz\over z(1-z)}  \nonumber\\
\int d^2 k_\bot \psi(z,k_\bot)
\int {d^{2} \kappa_\bot
\over
\kappa^{4}}\alpha_{S}
\cdot I_{\nu\mu}(\gamma^{*}\to V)
{\partial G(\bar{x},\kappa^{2})\over \partial \log \kappa^{2}}
\label{f1}
\end{eqnarray}
and are calculable in terms
of the gluon structure function of the target proton 
$G(x,\kappa^{2})$ taken
at $\bar{x}={1\over 2}x_{\Pom}=
{1\over 2}(Q^{2}+m_{V}^{2})/(Q^{2}+W^{2})$. Here $z, (1-z)$ and
${\bf k}$ are are the Sudakov lightcone variables of the $q$ and $\bar{q}$ 
in the vector meson. 
The typical
QCD hard scale is 
$\overline{Q}^{2}= z(1-z)Q^{2}+m^{2}$, where $m$ is the quark mass.
To the LL$\overline{Q}^{2}$, i.e., for $\kappa^{2} \lsim 
\overline{Q}^{2}$, one finds for pure $S$-wave mesons
\begin{eqnarray}
I^S_{0L}& =& -4QMz^2(1-z)^2{2 \kappa^2 \over \overline Q^4}
\left[ 1 + {(1-2z)^2\over 4z(1-z)} {2m \over M+2m}\right]\,, \nonumber\\
I^S_{\pm\pm}&= &
{2 \kappa^2 \over \overline Q^4}
\left[ m^2 + 2 k^2(z^2 + (1-z)^2) + {m\over M+2m} k^2 (1-2(1-2z)^2)\right]
\nonumber\\
I^S_{\mp\pm} &= &
4z(1-z)\Delta^{2}{k^2 \over 2\overline Q^4}
\left(1 + {6 \kappa^2 (1-2z)^2 \over \overline Q^2}\right)
\left[ 1 + {(1-2z)^2\over 4z(1-z)} {2m \over M+2m}\right]\nonumber\\
I^S_{0\pm} &=& - 2Mz(1-z)(1-2z)^2 \Delta{2 \kappa^2 \over \overline Q^4}
\left[ 1 + {(1-2z)^2\over 4z(1-z)} {2m \over M+2m}\right]\,,\nonumber\\
I^S_{\pm L} &=&-2 Qz(1-z)(1-2z)^2\Delta {2 \kappa^2 \over \overline Q^4}
{2 k^2 \over \overline Q^2}{M\over M+2m}\,,\label{h6}
\end{eqnarray}
where $M$ is an invariant mass of the $q\bar{q}$ pair. No separation of
the $S$ and $D$-wave has been done in \cite{IK}.

First, notice how the transverse and longitudinal Fermi motion of quarks in vector 
mesons are necessary for helicity flip and SCHNC,
which would suppress SCHNC in production of heavy quarkonia in
which quarks are nonrelativistic.
 Second, apart from
the double-helicity flip, $I^{S}_{\nu\mu} \propto \kappa^{2}$ which after
the $\kappa^{2}$ integration leads to
pQCD calculable
$A(\bar{x},Q^{2},{\bf \Delta}) \propto G(\bar{x},\sim {1\over 4}(Q^{2}+m_{V}^{2}))$.
Third, for the double-helicity flip
$A_{\mp\pm}(\bar{x},Q^{2},{\bf \Delta}) \propto 
G(\bar{x},\mu_{G}^{2})$ where $\mu_{G}^{2}\sim $(0.5-1)GeV$^{2}$ and it is not
pQCD calculable at any large $Q^{2}$. 
Finally, by exclusive-inclusive 
duality \cite{GNZlong} the above results for
$I^{S}_{\mu\nu}$ can be related 
SCHNC LT interference in diffractive
 DIS into continuum \cite{NZDIS97,NPZLT} and indeed the dominant 
SCHNC effect in vector meson production is the interference 
of production 
of longitudinal vector mesons by (SCHC) longitudinal and (SCHNC) 
transverse photons,
i.e., the element $r_{00}^{5}$ of the vector meson polarization 
density matrix. The overall agreement
between our theoretical estimates 
\cite{AINP} of the spin density matrix $r_{ik}^{n}$ for
diffractive production of the $\rho^{0}$ 
and the ZEUS \cite{ZEUS} and H1 \cite{H1} experimental data
is very good. More theoretical analysis of the sensitivity to
the wave function of vector mesons is called upon.

\section{Sensitivity of SCHNC to spin-orbit coupling in vector mesons}

Production of $D$-wave vector mesons nicely demonstrates a 
unique sensitivity of helicity flip in $\gamma^{*}p\to Vp'$
to spin-orbit coupling \cite{KNZ98,IN99}:
\begin{eqnarray}
I^D_{0L}&=&-{Q \over M}\cdot
{32 {\bf r}^4 \over 15 (M^2 + Q^2)^2}\cdot
\left( 1 - 8 {M^2 \over M^2 + Q^2}\right) {\bf \kappa}^2 \, ,\nonumber\\
I^D_{\pm\pm}&=&
{32 {\bf r}^4 \over 15 (M^2 + Q^2)^2}
\cdot  \left( 15 + 4 {M^2 \over M^2 + Q^2}\right) {\bf \kappa}^2 \, ,\nonumber\\
I^D_{\pm L}&=&
 - {24 \Delta  Q  \over M^2 + Q^2}{32 {\bf r}^4 \over 15 (M^2 + Q^2)^2}\cdot
{24 Q  \over M^2 + Q^2} {\bf \kappa}^2 \, ,\nonumber\\
I^{D}_{L\pm}&=& {8 \Delta \over M}{32 {\bf r}^4 \over 15 (M^2 + Q^2)^2}\cdot
 \left( 1 + 3 {M^2 \over M^2 + Q^2}\right)  {\bf \kappa}^2 \,,\nonumber\\
I^{D}_{\pm\mp}&=&
 \Delta^{2} 
\cdot { 32 {\bf r}^4 \over 15 (M^2 +Q^2)^2}\cdot
\left( 1 - {96 \over 7} {{\bf \kappa}^2 
{\bf r}^2 \over M^2 (M^2 + Q^2)}\right),
\end{eqnarray}
where $4r^{2}=M^{2}-4m^{2}$.  In the $D$-wave state the total spin 
of $q\bar{q}$ pair is predominantly opposite to the spin of the $D$-wave 
vector meson. As a results, in contrast to
$S$-wave states there is no nonrelativistic suppression of helicity
flip. Such an enhancement of SCHNC may facilitate the $D$-wave 
vs. $2S$-wave assignment of the $\rho'(1480)$ 
and $\rho'(1700)$ and of the $\omega'(1420)$ and $\omega'(1600)$,
which remains one of hot issues in the spectroscopy of vector mesons.
Notice abnormally large higher twist corrections. For instance, 
$A_{0L}$, and $LT$ interference thereof, 
changes the sign at  $Q^2 \sim 7 m_{V}^{2}$. The ratio 
$R^{D}=\sigma_{L}/\sigma_{T}$ has thus a non-monotonous $Q^2$ behavior 
and $R^{D} \ll R^{S}$. Furthermore, $R^{D} \lsim 1$ in a broad range of 
$Q^{2} \lsim 225 m_{V}^{2}$. \\
 
\section{SCHNC physics at Jefferson Lab}

Above we focused on SCHNC in QCD pomeron exchange which is described in
pQCD by a generalized two-gluon ladder in the $t$-channel. There will be  
a substantial secondary reggeon contribution to diffractive $\rho^{0}$
and $\omega^{0}$ production at CEBAF energies even after energy upgrade.
In pQCD the reggeon exchange is modeled by generalized quark-antiquark
ladder in the $t$-channel \cite{Reggeon}. An exhaustive analysis of SCHNC 
for secondary
reggeons has not been carried out yet. Still, the above predictions of
SCHNC are fully applicable to, and can be tested in, the hidden-strangeness 
$\phi^{0}$ and radial and angular-excited $\phi'$ diffractive 
photo- and electroproduction at CEBAF, because
secondary reggeons do not contribute to the $\phi^{0}$ production.
Recall, for instance, the Zweig rule.  

\section{Dramatic impact of SCHNC diffraction upon the small-$x$
behaviour of transverse spin structure function $g_{2}$}
\begin{figure}
\epsfysize 1.3 in 
\epsfbox{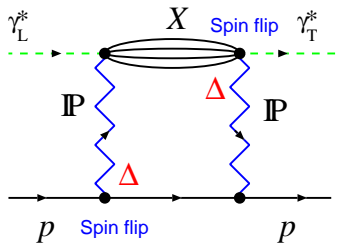}
\caption{The unitarity diagram for diffractive contribution to
$g_{LT}$.}
\label{fig:2}
\end{figure}

The transverse spin asymmetry in polarized DIS is proportional to the
amplitude of forward CS 
$\gamma^{*}_{L}p\!\!\uparrow \to \gamma^{*}_{T}p\!\!\downarrow$. 
 This CS 
amplitude and the transverse spin asymmetry 
are proportional to $g_{LT}=g_{1}+g_{2}$. Because 
the photon helicity flip is compensated by the 
target proton helicity flip, the familiar forward zero of this helicity 
amplitude is lifted. However, in the standard two-gluon $t$-channel 
tower approximation the cross-talk
of the target and beam helicity flip  is only possible if 
in the Gribov-Lipatov decomposition of the gluon propagator
only one of the gluons is having the nonsense polarization
whereas the second one has the transverse polarization. The price 
one pays for such a combination of polarizations of gluons is the suppression of
small-$x$ behaviour of $g_{LT}$  by the
extra factor $\sim x$ compared to the pomeron exchange in which 
the both gluons have the nonsense polarization .

The more familiar argument for the vanishing 
$A_{2}$ has been the parton model Wandzura-Wilczek relation between 
$g_{LT}$ and $g_{1}$ \cite{WW}:
$$
g_{LT}(x,Q^{2})=\int_{x}^{1} {dy \over y} g_{1}(x,Q^{2})\, ,
$$
which entails that $A_{2}\sim A_{1}$.    
Because the diffractive pomeron exchange does not contribute to $g_{1}$, 
the WW relation can be reinterpreted as a vanishing pomeron exchange
contribution to $g_{LT}$. This vanishing of the pomeron contribution 
has been the principal motivation behind
the much discussed Burkhardt-Cottingham \cite{BC} (BC) sum rule
$\int_{0}^{1} dx g_{2}(x,Q^{2})=0$. Our recent discovery is that diffractive 
SCHNC destroys the both WW relation and BC sum rule \cite{G2}.

The opening of diffractive DIS channels affects, via unitarity, the
Compton scattering amplitudes. In \cite{G2} we have shown how diffractive 
LT interference in conjunction with  spin-flip pomeron-nucleon coupling 
$r_{5}$ give rise to the 
transverse spin asymmetry $A_{2} \propto x^{2}g_{LT}$ which does not vanish 
at small $x$. The building blocks of the  unitarity diagram shown in fig.~2 
are the diffractive amplitudes $\gamma^{*}p \to p'X$ in which there is
a helicity flip sequence, $\gamma^{*}_L \to X_{L} \to \gamma^{*}_T$ in the top 
blob and helicity flip sequences either 
$p\!\!\uparrow \to p'\!\!\uparrow \to p\!\!\downarrow$ 
or $p\!\!\uparrow \to p'\!\!\downarrow \to p\!\!\downarrow$
in the bottom blob. The both amplitudes are proportional to ${\bf \Delta}$
and vanish in the forward direction, but upon the integration 
over the phase space of $p'X$ one finds the nonvanishing 
$\int d^{2}{\bf \Delta} \Delta_{i}\Delta_{k}$ and unitarity driven transition 
$\gamma^{*}_{L}p\!\!\uparrow \to \gamma^{*}_{T}p\!\!\downarrow$ which does
not vanish in the forward direction. In the old
Regge theory language it can be reinterpreted as the two-pomeron cut
contribution which for the QCD pomeron has about the same small-$x$ 
behaviour as the pomeron exchange. The principal difference from the
single pomeron exchange is that the unitarity diagram starts with the 
four-gluon state in the $t$-channel and four gluons can furnish the
cross-talk of the beam and target helicity flip with pure nonsense
polarizations of all the four exchanged gluons.

Our result \cite{G2} for the diffraction-driven $g_{LT}$ reads 
\begin{equation}
g_{LT}(x,Q^{2}) \propto {1\over x^{2}} r_{5}\int_{x}^{1} {d\beta\over \beta}
g_{LT}^{D}(x_{\Pom}={x\over \beta},Q^{2})\, .
\end{equation}
It rises steeply at small $x$. It is the scaling function of $Q^{2}$ because 
the diffractive LT structure function $g_{LT}^D(x_{\Pom},Q^{2})$ is the
scaling one. The corresponding 
transverse spin asymmetry $A_{2} \propto xg_{LT}/F_{1}$ does not vanish 
at small $x$, furthermore, at a moderately small $x$ it even rises because
$g_{LT}^D(x_{\Pom},Q^{2}) \propto G^{2}(x_{\Pom},\overline{Q}^{2})$ where
$\overline{Q}^{2} \sim$ 0.5-1 GeV$^2$. 

\begin{figure}
\vspace{9pt}
\epsfysize 1.8 in 
\epsfbox{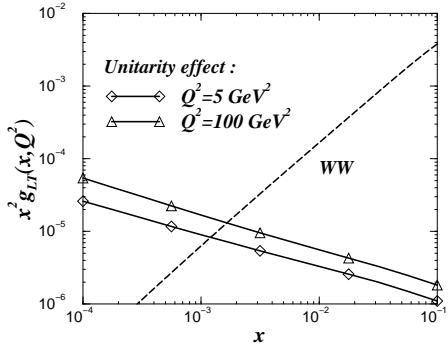}
\caption{The unitarity correction to, and WW relation based 
evaluation of, $g_{LT}$
\protect\cite{G2}.}
\label{fig:3}
\end{figure}

In fig.~3 we show how the steeply rising unitarity 
correction overtakes at small $x$ the standard  $g_{LT}$ evaluated from the 
Wandzura-Wilczek 
(WW) relation starting with fits to the world data on $g_{1}$. 
As such our unitarity
effect is the first nontrivial scaling departure from the WW relation

Finally, the above breaking of the WW relation implies $g_{LT} \gg g_{1}$ 
and $g_{2} = g_{LT}$ at very small $x$. Consequently, the unitarity-driven
rise of $g_{2}$ 
destroys the BC sum rule because the BC integral would diverge severely.
Incidentally, the BC sum rule has always been suspect.

\section*{Conclusions}

The QCD pomeron exchange does not conserve the $s$-channel helicity.
The mechanism of SCHNC is well understood. SCHNC offers an unique 
window at the spin-orbit coupling in vector mesons. SCHNC in diffractive
DIS drives, via unitarity relation, a dramatic small-$x$ rise of the transverse
spin structure function $g_2$ which breaks the Wandzura-Wilczek relation
and invalidates the Burkhardt-Cottingham sum rule. Jlab is perfectly posed
to study SCHNC QCD pomeron exchange in  photo- and
electroproduction of the $\phi$ and its radial and orbital excitations.\\

\noindent
{\bf Acknowledgments.} Thanks are due to C.Carlson and A.Radyushkin for the
invitation to the Workshop.

\end{document}